# High-Resolution Mid-Infrared Observations of Low-Luminosity AGN with the Keck Telescope and LWS

Bruce Grossan[1]

[1]Lawrence Berkeley National Laboratory Institute for Nuclear and Particle Astrophysics (INPA); Bruce_Grossan@lbl.gov

## ABSTRACT

We present high-resolution Keck LWS SiC filter (10.5-12.9 µm) mid-IR (MIR) images and photometry of a sample of nearby LLAGN. All MIR images show extended emission on small scales (e.g. 90% maximum widths 18 - 106 pc) deep within the nucleus. Because the profile of this small-scale emission is inconsistent with the near-IR stellar emission profile, we conclude the MIR emission is not from the stellar population. Because strong starburst emission is not reported for our objects, we eliminate this origin as well, and conclude that the extended component of the MIR emission is most likely due to a warm dust structure deep in the nucleus, consistent with the gross details of a Unified Model Torus.

Both MIR and X-ray flux measurements are highly insensitive to the effects of pure dust extinction; therefore the ratio of MIR and X emission is discriminator of emission processes that is nearly insensitive to dust extinction. X and MIR emission are known to be correlated in quasars and Seyfert galaxies. Our LLAGN objects follow this correlation with small scatter down to extremely low luminosity, suggesting that the same continuum emission mechanisms are present in LLAGN as in Seyferts and quasars. This result suggests that weak UV emission relative to other bands in LLAGN compared to that in Seyferts and quasars could be due to dust extinction; it then follows that modifications to standard AGN continuum emission models for LLAGN (e.g. ADAF accretion) are not required.

## 1. INTRODUCTION

### 1.1. The Nature of LLAGN Emission - A Seyfert Continuum?

Low-Luminosity Active Galactic Nuclei (LLAGN) have been variously defined in terms of their Hα luminosity (e.g. see Ho, Filippenko & Sargent 1997), or for Seyfert-like objects, as lower in luminosity than NGC4051. Spectroscopically, the term typically encompasses LINER galaxies and objects in between LINERs and Seyferts. The observed spectral lines tend to be of lower ionization states than in bona fide Seyfert nuclei, but with more line activity characteristic of non-stellar and non-thermal processes than from starbursts. Various mixes of mechanisms have been invoked to explain LLAGN emission (see e.g. Ho, Filippenko & Sargent 1993). However, the presence of massive black holes in the nucleus of nearly every galaxy with a spheroidal component (e.g. Gebhardt et al. 2000), and the observation of LLAGN lines typically produced by photoionization by a hard, non-thermal continuum leaves little doubt of some connection to classical AGN accretion-powered processes. What is unclear is the degree of modification of the classical AGN emission processes, and how they are linked to luminosity.

Ho (1999) studied a sample of LLAGN, and concluded that there is a systematic lack of UV emission compared to AGN. He further proposed that the weak UV emission of LLAGN pointed to an intrinsic difference between AGN and LLAGN spectral energy distributions (SEDs), and that this intrinsic SED difference was explained by advection-dominated accretion flow (ADAF) . In the ADAF model, the accretion disk is "stood-off" from the black hole some large distance, e.g. 100 Schwarzschild radii (see e.g. Rees et al. 1982), to produce low accretion and radiative efficiency as required for the low luminosity in these objects. Ho (1999) did explicitly state, however, that the data do not rule out the possibility that extinction, rather than intrinsic physics, might explaining the lack of UV emission.

Grossan et al. (2001a) described high-resolution N band observations of the LLAGN M81 with the MIRLIN camera that effectively limited the flux measurement to the nucleus, whereas previous, larger aperture measurements did not. The paper reported that the measured flux was 8 times that predicted by Quataert et. al. (1999) fit to an ADAF model. However, a new central black hole mass reported by Devereaux et al. (2003) now makes it is possible to fit the M81 SED with an ADAF model; therefore M81 is *not* a "refuted ADAF".

Additional recent literature on the nature of LLAGN emission includes Ho et al. (2001), which reports on a study of X-ray images of nearby Galaxies with the Chandra Observatory. The study finds that LLAGN type 1 (broad emission lines) follow the same correlation of H-alpha Luminosity vs. X-ray Luminosity as Seyferts, but that type 2 (narrow line) objects have more complicated behavior. Transition (LINER/HII) objects, the paper concluded, are unrelated to AGN. Many type 2 LINER and Seyfert objects have dominant X-ray point-sources, as expected for AGN emission; however, these objects are clearly X-ray under-luminous compared to Type 1 objects at the same Hα luminosity. While this is given as tentative evidence that the Unified Model does not hold for type 2 objects at the lowest luminosities, the authors stated that the presence of a highly absorbed component ($N_H >\sim 10^{25}$ cm$^{-2}$) or a confused contribution from an X-ray binary cannot be refuted from their data. For type 1 objects, these results support the general picture of a Seyfert-like accretion-powered continuum photoionizing gas clouds; gas composition, environment, etc. will determine if the object has more Seyfert or LINER characteristics. (We note that the broad line objects in the Ho et al. (2001) sample do not extend to the lowest luminosities, with just 2 type 1 objects in 18 at $L_{x,2\text{-}10\,keV} < 10^{39.5}$ erg s$^{-1}$. The behavior of type 1 objects at the lowest $L_{x,2\text{-}10\,keV}$ is not well described by this paper, therefore, in this luminosity range, the behavior of the type 2 objects is only given relative to the *extrapolation* of high luminosity Seyfert galaxies.)

Contrarily, Ptak et al. (1998) have reported that LLAGN, including type 1 LINERS, have different variability behavior than for Seyferts, and therefore may have different intrinsic emission mechanisms. The work concluded that the data supported ADAF accretion in LLAGN.

## 1.2. LLAGN Observation Program
### 1.2.1 LLAGN Emission and the AGN MIR/X Relation

For the past few years we have been studying the behavior of LLAGN in the mid-IR (MIR). At MIR wavelengths, dust extinction is extremely small. At the same time, warm dust (~ 100 K) radiates copiously in the MIR bands. MIR observations therefore permit a special window onto warm dust structures in dusty environments. There is also no extinction due to pure dust at hard X-ray wavelengths; though gas does attenuate X-ray emission, spectroscopic measurements allow a fit to the absorption and a subsequent flux correction. The ratio of MIR and X emission is therefore an especially valuable tool to compare dusty and non-dusty instances of the same objects. Krabbe, Börker, Maiolino (2001; KBM hereafter) showed that in Seyferts and quasars, there is an excellent correlation of $L_{MIR}$ and $L_X$ over several decades of luminosity. The relation of $L_{MIR}$ and $L_X$ therefore seems to be an excellent "dust-immune" discriminator of bona fide Seyfert activity.

In order to further investigate LLAGN with this discriminator, we selected a number of LLAGN with published X-ray measurements as targets for MIR measurements. In Grossan et al. (2001b) we showed that the ratio of $L_{MIR}/L_X$ for M81 was close to the KBM AGN relation; in Grossan et al. (2001a) we showed preliminary evidence (large uncertainties due to weather) that M58, M81, and M87, all LLAGN with either LINER or LINER/marginal Seyfert classifications, also had $L_{MIR}/L_X$ ratios consistent with the KBM relation.

### 1.2.2 LLAGN Nuclear Structure

MIR observations have an additional special property: because the seeing in the MIR is superior to that at shorter wavelengths, the largest telescopes can resolve exceptionally small scales in the most nearby AGN, sufficient to probe structure deep within the nucleus. At Keck, we achieved a FWHM resolution of 0.34-0.38", corresponding to about 6.3 pc at the distance of M81. In Grossan et al. (2001b) we showed evidence of extended emission around the nucleus at N band in M81. This emission (reported as 70-120 pc in size) is grossly consistent with modern theories of Unified Model dust tori: (i) broad-band MIR emission, consistent with emission from an optically thick region, was detected with a profile and an intensity inconsistent with stellar emission, and (ii) the size scale of the emission is consistent with some theoretical predictions (i.e. Fadda et al. 1998).

Because this emission can only be resolved by the largest telescopes in the most nearby LLAGN, we have restricted our study sample (and the discussion in this paper) to LINER or marginal Seyfert/LINER classified objects that are within 17 Mpc. In order to clarify the nature of our findings at Palomar (Grossan 2001a; Grossan 2001b) we obtained Keck time for the study of our LLAGN sample. In the remainder of this paper we report on MIR observations of M81/NGC3031, NGC 4203, M104/NGC4594/Sombrero, and M94/NGC4736 from our sample using the Keck Telescope and LWS camera, along with measurements at N band with the MIRLIN instrument at the Palomar 200" telescope (Grossan et al. 2001a).

## 2. OBSERVATIONS

The Keck LWS instrument, described elsewhere (Jones & Puetter 1993), has a 256X256 MIR sensitive detector and produces images or spectra throughout the MIR (3.5 – 25 μm). The pixel size is 0.08"/pix. The instrument subtracts background by means of a chopping secondary, with data taken in a four position chop/nod pattern.

We report aperture photometry of our targets. The flux given is the integrated light in all pixels whose centers are within the indicated diameter minus a local background estimated by sampling an annulus with a large inner diameter. The error in the background is estimated via the standard deviation of the pixels in the background annulus. No corrections are made for flux in pixels partially within the given diameters (because this error is small for our radii).

Two nights at Keck are reported, UT 2002/4/21 and UT 2003/3/17. The observations are described in Table 2a. (The original program that we proposed to Keck included spectroscopy, but instrument problems precluded these measurements.) Unfortunately, because of both weather and instrument problem impact on our schedule, we were unable to make any photometric measurements in bands other than SiC, and we observed only two standard stars, HR5315 and HR4954. The SiC filter is a relatively broad band filter covering from 10.5-12.9 μm at ≥ 80% transmission.

The standards in the 2002 run showed more than 0.6 mag variation, so measurements were not photometric. Repeated measurements of the same standards during the 2003 run showed good stability (0.01 mag, far less than a reasonable systematic error estimate). The 2003 calibration values derived separately for the two standards differed by only 0.012 mag. The calibration factors showed no correlation with airmass, and the target airmass values were within the range of standard star airmass during observations, therefore no extinction corrections were applied. Results are given in Table 3. The statistical error in the calibration was dominated by the variance in calibration measurements taken at different times during the night; this variance is most likely due to time-varying transparency. This error was added in quadrature with the statistical errors in the photometric measurements. No systematic error estimate is applied. Additional measurements from Palomar are given in Table 2b.

## 3. RESULTS

### 3.1 Imaging Results

Figure 1 gives contour plots of our objects observed with the Keck LWS. Most objects show dramatic extended emission of the nucleus. The profiles of these galaxies along both the major and minor axes are shown in Fig. 2, with an averaged standard star profile for comparison. All nuclei are significantly more extended than the standard star profiles, and most are obviously asymmetric (elliptical). These results confirm those reported from Palomar with the MIRLIN instrument (Grossan et al. 2001b), but the sensitivity in these images is far greater. Table 2 gives the FWHM and 90% diameters of all objects, both in angular size and size at the source.

In order to test if the nuclear structures are related to the large-scale galactic structure, we show the orientation of the optical structure (the PA from the RC3 catalog) and the orientation of a non-rigorously chosen MIR major axis orientation on the contour plots. (The MIR major axes were chosen by examining the contours determined by our software and averaging the orientations of the extrema of each contour outside of 1 FWHM and well above the noise contours; the results are non-unique and highly uncertain. The major axes are used only for the purpose of searching for gross relationships between the MIR and optical/NIR structure.) No obvious relation is apparent.

Figure 3 shows cuts along the MIR major and minor axes from both Keck SiC and HST 1.6 μm images from the MAST archive of NICMOS images (where available). The cuts are scaled to the maximum pixel value in each cut. The MIR and NIR profiles are very different in all cases.

### 3.2. Flux measurements

Tables 2b and 2c give aperture photometry measurements for diameters yielding 68%, 90%, and 95% of our standard star maximum fluxes. In addition, we provide an "Extended Flux

Ratio", the ratio of the aperture photometry to that for a PSF with the same 68% diameter flux, as a measure of extended flux.

### 3.3. X-ray vs. MIR Emission Plots

Figures 4 a and 4b show a plot of 2-10 keV X-ray vs. MIR flux and luminosity for our sample. These LLAGN are very close to the trend for Seyferts and quasars reported by KBM, certainly well differentiated from the locus for starburst galaxies. M94 and NGC 4203 are outliers, however, both about a factor of 3 too faint in X or bright in MIR. These objects are the two lowest in X-ray luminosity, though not in X-ray flux.

KBM emphasized the importance of applying reddening corrections to their sample; some of their type 2 objects had extraordinarily high $A_v$ up to 60 mag. We corrected our LLAGN for known reddening (see Table 1) assuming $A_{SiC} = A_N = A_{10.6\,\mu m}$, and $A_{10.6\,\mu m} / A_V = 0.05$ (Rieke & Lebofsky 1985) and assuming Milky Way-like extinction, i.e. $A_V/E(B-V) = 3.1$. Note that any nuclei with $A_V < 2$ mag ($E(B-V) \leq 0.6$) have $< 10\%$ extinction in the MIR. None of our sample objects have any indicators of reddening which would yield significant corrections with the exception of M58, for which a correction as large as 38% (E(B-V) up to 2.26 mag) was found in the literature.

## 4. DISCUSSION

### 4.1.1 Extended Emission

An inspection of Table Ib shows that the profiles of all the objects we observed with Keck are extended. The most likely origin of extended emission in the MIR is from warm dust. At the small length scales that our images show, it is tempting to associate this emission with the proposed Unified Model obscuring torus. Dusty starburst regions can also produce luminous MIR emission, but there is no spectroscopic evidence of strong starburst activity reported near the nuclei of these galaxies. Another possible source of MIR emission might be that from photospheric and circumstellar dust in the stellar population. Such emission would be expected to follow the stellar emission profile given in H or K bands. Fig. 3 clearly shows that the observed major and minor axis MIR profiles are inconsistent with profiles along the same axes in HST F160 band (essentially H band) images, where starlight dominates. All our objects have de Vaucouleurs stellar profiles extending to large radii (i.e. throughout the bulge), and which are very smooth to the limit of the HST image resolution. The MIR extended emission is therefore not related to the stellar emission component. In fact, this component is expected to contribute only negligibly in our small apertures. In Grossan et al. (2001b) the stellar contribution to the N band measurement of M81 was estimated to be only 11% in the 3.9" aperture. (Using the same procedure, we estimate the stellar contribution to M81 N band emission to be 5%, 8%, and 10% for the 0.8", 1.76", and 2.72" apertures, respectively). In the overall sample, we have shown that extended MIR emission within the nucleus is common in LLAGN.  These findings are therefore consistent with a Unified Model Torus or similar warm dust structure in LLAGN in our sample, though we do not image an obvious torus separate from the central continuum source (i.e. the accretion disk region).

We expect that MIR spectroscopy should tell us about the composition of the MIR nuclear emission via dust emission and absorption features.  Unfortunately, atmospheric absorption bands make observing these features from the ground challenging.  It is difficult to observe them from space as well; because of the limited resolution of current space-born instruments, including the Spitzer Space Telescope, nuclear region emission cannot be isolated, and dust signatures could come from dust far from the nucleus.

### 4.1.2 Extended MIR Emission in Classical AGN

KBM observed 3 close type 2 Seyferts and reported nuclear extended emission from one object: Circinus, d=4.0 Mpc, with extended emission detected to ~4-5 arc sec or ~80-100 pc radius. Though the apparent presence of extended emission in Circinus may seem encouraging for additional studies of tori, it is also a concern that the other two objects did not apparently have extended emission.  Because of the limited resolution in their MANIAC images, along with the brighter nuclei in their Seyfert sample, and because of the added complexity of high extinction and starburst concentrations in these systems, the non-detections most likely do not amount to sensitive limits on torus emission.

### 4.2 X-MIR Correlation

The KBM correlation clearly extends to lower luminosity and flux with this LLAGN sample.  This correlation suggests that LINER and Seyfert/LINER transition objects in our sample are dominated by emission components in the MIR and X-ray bands that are physically similar to those in Seyferts and quasars.  These results argue against LLAGN accretion processes with different SEDs than in Seyferts and quasars, at least in the MIR and X bands.

Should one expect this correlation to continue to arbitrarily low luminosity? Consider a relationship between MIR and X emission which is independent of luminosity.  Moving toward lower and lower luminosities, eventually some other source of MIR emission should dominate the MIR emission from the near-nuclear region, e.g. from the atmosphere of evolved stars, or if present, from star formation regions embedded in dust.  Note however, that star formation has distinctive spectral features, and is typically localized in knots.  Any star formation emission that would dominate the MIR emission should therefore be recognizable with high-resolution imaging (if not within ~ 1 FWHM of the nucleus) and spectra.   In X-rays, emission from accreting binary systems would likely dominate the nuclear emission.  Note, however, that these systems have characteristic X-ray spectra and are also not extended; again, any such activity that would dominate the X-ray emission should be recognizable by high resolution imaging unless within ~ 1FWHM of the nucleus or by spectra. These likely circumstances are therefore distinguishable from AGN emission given sufficient resolution and/or sufficient quality spectra. All nuclei that have detectable AGN/LLAGN emission should, by these simple arguments, be near the KBM line. The two objects in this paper off the KBM relation have X-ray spectra consistent with LINER objects and have no reported strong starburst optical/NIR spectral features. Because of the significant spread in the KBM relation, these two discrepant objects are not statistically significant. It would be very informative to examine, however, ~ 10 objects below $L_X = 10^{40.1}$ erg s$^{-1}$ to look for a breakdown in the relation. Substantial X-ray variability is common, and could be the cause of these stray data, though factor 3 X-ray variability in LINERs is uncommon (Ptak 1998).

We believe that our work documenting microstructure within the nuclei of nearby galaxies shows great promise for directly verifying an essential feature of the Unified Model, the AGN torus. Further, this work shows the great utility in increases in high-resolution MIR instrumentation. In particular, we look forward to the increased sensitivity of the NGST, which will yield more precise profile measurements as well as extending the sample to even lower luminosity. The NGST will also have the capability to map dust emission from the putative torus spectroscopically without the difficulties of atmospheric absorption features that plague such work from the ground.

## 5. CONCLUSIONS AND SUMMARY

High-resolution MIR imaging of 4 nearby LLAGN, LINERS and transition LINER/Seyferts, shows extended MIR emission deep within the nucleus. The profile of this emission is inconsistent with the NIR stellar emission profile. Because this emission is bright in the MIR and not related to stellar emission, it is consistent with an origin in a Unified Model Torus.

We showed that our LLAGN follow the KBM relation for Seyferts to very low luminosity, with the possible exception of the two lowest X-ray luminosity objects. Ignoring the two outliers, these data strongly suggest that the same physical processes of continuum emission must be operating in the nuclei of LLAGN and Seyferts and quasars. The chief differences remain in line emission and jets and related activity (though we note that even the very low-luminosity M87 has substantial jet activity).

We thank the American Astronomical Society for their support of this work through their award of a Small Research Grant to Bruce Grossan. We also thank the Institute for Nuclear and Particle Astrophysics and the Institut d'Astrophysique Spatiale, including support staff, for institutional support of Dr. Grossan in the course of this work. We thank the staff of Keck and Palomar observatories for providing support for the observations described herein, especially

Randy Campbell for his help with LWS data reduction. We thank the MIRLIN team, Drs. Varoujan Gorjian, Michael Werner, Michael Ressler, and Luisa Rebull, for their contributions to the acquisition and analysis of the MIRLIN data. This work made use of the NASA Astrophysical Data System (ADS), the NASA/IPAC Extragalactic Database (NED), and the Multi-Mission Archive at the Space Telescope Institute (MAST).

# APPENDIX - CONVERTING FROM SIC FLUX MEASUREMENTS TO N FLUXES

At Keck the LWS camera has such a high flux that a normal N filter cannot be used without saturating the chip. A SiC filter is therefore provided. Below, we derive a simple approximate conversion factor between the two bands.

It is common practice in the literature to give a flux density in mJy at the "center" of a filter bandpass derived from broadband photometry. Typically, including in Krabbe, Borker, Maiolino (2001), magnitudes are converted to mJy by simply adopting a definition for 0.0 mag without making a spectral correction, i.e.

(1) $F_\lambda = 10^{-m_\lambda/2.5} \times F_{0\lambda}$

where, for $m_\lambda$ = magnitude in the N filter, $F_{0\lambda} = F_{0N}$ is taken to be $4.26 \times 10^{-2}$ Jy at $\lambda_0 = 10.2$ μm.

Converting between magnitudes in the N and SiC bands is more problematic, because the spectrum of the target and standard must be considered. The spectra of LLAGN in the MIR are very poorly known, in fact, even the spectra of AGN are poorly known in this band. ISO recorded high-quality spectra of a large sample of AGN (e.g. Clavel et al. 2000). However, ISO's poor angular resolution (4-5") permits severe pollution by galaxy emission. In fact, these spectra all show obvious features of galaxy dust emission, and limited observations are available to guide our understanding of the intrinsic contribution of these features. Clavel et al. 2000 fit a power law plus dust features to the ISO AGN spectra, and found the slope to be $-0.84 \pm 0.24$ for objects with some evidence of broad lines. (For strictly narrow line Seyferts, a slope of $-0.82 \pm 0.37$ was reported, consistent with that for broad lined objects. ) From this point on, we assume our targets are well approximated with power-law spectra in the MIR, and our standard stars have featureless black body spectra.

For purely power law objects and purely black body standards, the ratio of target and standard signals from a photon-counting device is:

(2)     $S_{pow}/S_{BB} = \int d\lambda\, F_{\lambda,powl} T_{\lambda filter}\, T_{\lambda atm}\, /\, \int d\lambda\, F_{\lambda,BB} T_{\lambda filter}\, T_{\lambda atm}\, \lambda$

where we assume the response of the camera system is flat in this spectral region, and $F_{\lambda,powl}$ is the flux density of a power law, $F_{\lambda,BB}$ is the flux density of a black body spectrum, and $T_{\lambda filter}$, $T_{\lambda atm}$ are the transmission functions of the filter and the atmosphere, respectively.

The SiC transmission curve is provided by the Keck Observatory. The N filter transmission curve for OSCIR is readily available, and probably representative of both the N filter used with MANIAC in Krabbe, Borker & Maiolino (2001), and with that used in MIRLIN. The N filter has quite sharp cutoffs at 8.1 and 13.4 μm, with some odd "emission"-like features between 8.25 and 8.75 μm. The transmission rises steadily from 0.67 to 0.92 in between the cutoffs. The SiC filter is quite different, with a response that grows from 10-70% between 10.25 and 11 μm, a nearly symmetric, broad peak from 70-80% transmission between 11 and 12.8 μm , and steep drop from 12.8 - 13.1 μm. These transmission curves are plotted in figure A-1.

The standards used in the SiC observations are K2 and K5 stars, with $T_{eff}$ = 4370, 3920 K respectively.

Using the data above, we can calculate a correction between SiC and N measurements, for an adopted $T_{eff} = 4145$ K and the given power law thusly:

(3)    $R_{NSiC} = (S_{pow}/S_{BB})_N / (S_{pow}/S_{BB})_{SiC} = 0.943 \pm 0.005$

where the uncertainty reflects only $\pm 1\ \sigma$ values of the spectral slope. The value is extremely insensitive to $T_{eff}$; the value for an A0 star $T_{eff} = 10,800$ is 0.3% smaller.

For our objects and standards, we convert to an approximate flux density similar to that given in Krabbe, Borker, Maiolino (2001) by:

(4)    $F_{N,eff} = 10^{-Nstd/2.5}\ F_{0\lambda}\ R_{NSiC}\ (S_{target}/S_{std})_{SiC}$

where $N_{std}$ is the given magnitude of the standard star, $S_{target}$ and $S_{std}$ are the integrated countrates of the target and standard stars.

If there are strong spectral features in the intrinsic spectrum of these objects outside of the SiC filter bandpass, but inside the N bandpass, this approximation will result in significant errors.

# TABLES

Table 1 Sample Description and Observations

| Target | Class | Distance (Mpc) | Nuclear Reddening $E(B-V)$ (mag) [Reference] | Nuclear Reddening Correction at N | Palomar / MIRLIN 2001 | Keck 2002 | Keck 2003 |
|---|---|---|---|---|---|---|---|
| M81 N3031 | "LINER w/ marginal SY1.5" | 3.6 [F] | ~0.1 [1] | 1.014 | | X [Non-photometric Imaging] | |
| NGC4203 | L1.9 [H01] | 15.1 [H01] | <~ 0.3 [B] | 1.04 | | | X |
| M87 NGC4486 | L2 [H] | 16.8 [H] | (A_V = 0.12) [H] | 1.017 | X | | |
| M58 NCG4579 | S1.9/L1.9 [H] | 16.8 [H01] | 0.16 gal + 0.21-2.1 [H] | 1.054 | X | | |
| M104[3] NGC4594 Sombrero | L2 [H01] | 9.8 [H01] | 0.2gal+0.25-0.28 internal; decrement and X-ray | 1.07 | | | X |
| M94[3] NGC4736 | L2 | 4.3 | gal 0.018 Eracleous | 1.003 | | | X |

    In general, authoritative reddening measurements are not available for the nuclei of these galaxies. Typically, one must choose between values derived from a Balmer decrement or X-ray column. The former is dependent on an assumed intrinsic value of the Balmer decrement, and indicates the reddening to the narrow or broad emission line regions, not the nucleus. The latter indicates the gas column to the nucleus, but assumes a gas/dust ratio. The Balmer decrements are usually higher (see Ho 99, for example).

    [1] Kong et al. 2000 - Reddening map makes location of nucleus unclear. Text compares to Filippenko & Sargent (1988) E(B-V) = 0.092, determined by narrow line Balmer decrement, and says the value is consistent. ; Ho reports galactic = 0.12; internal = 1.4 by Balmer decrement, < 0.2 by X-ray column.

    [H] Ho 1999 For M87 Note that Ho gives 08 gal +1.0 mag internal from the Balmer decrement, but in the same paper argues that strong UV emission demonstrates negligible extinction and that spectral modeling suggests A_V = 0.12.

    [H01] = Ho et al. 2001

    [F] Freedman et al. 1994

    [K] Tsvetanov

    [B] Barth et al. 1998 give a Balmer decrement of 1.03 and show that the nucleus is UV bright, a direct demonstration of low extinction to the true nuclear source. The low extinction to the nucleus argues that Balmer decrement, among their sample, correlates with UV detectability. They report this population

    [R] M. Serote Roos et al. MNRAS 301,1 1998

    Note in M58 the Balmer decrement is much larger than the X-ray column.

Table 2a Keck 2003 Imaging Results Summary

| | | Aperture Diameter[1] SiC (mag) | | | Aperture Diameter[1] Flux (mJy) @10.2 μm[2] [Extended Flux Ratio][3] | | | |
|---|---|---|---|---|---|---|---|---|
| Target (spectral class) | Log Lx (erg s$^{-1}$ cm$^{-2}$, 2-10 keV) | 0.8" 10 pix (68% PSFD) | 1.76" 22pix (90% PSFD) | 2.72" 34 pix (95% PSFD) | 0.8" 10 pix (68% PSFD) | 1.76" 22pix (90% PSFD) | 2.72" 34 pix (95% PSFD) | Time on-source (total) |
| M81 N3031 "LINER w/ marginal SY1.5" | 40.2 [H01] | ** | ** | ** | **1.0 ± 1.8 [1.0] | **1.81 ± 0.04 [1.37]] | **2.17 ± 0.04 [1.55] | |
| NGC4203 L1.9 | 40.08 [H01] | 8.43 ± 0.05 | 7.87 ± 0.06 | 7.73 ± 0.08 | 18.1 ± 0.8 [1.0] | 30.3± 1.6 [1.26] | 34.6 ± 2.5 [1.37] | |
| M104[3], N4594 L2 | 40.14 [H01] | 8.96 ± 0.08 | 7.96 ± 0.07 | 7.51 ± 0.07::(3) | 11.1 ± 0.8 [1.0] | 28.0 ± 1.8 [1.90] | 40.3 ± 2.7::(4) [2.72] | |
| M94[3], N4736 L2 | 39.36 [R01] | 8.55 ± 0.04 | 7.34 ± 0.03 | 6.77 ± 0.02 | 16.3 ± 0.5 [1.0] | 49.4 ± 1.2 [2.29] | 83.8 ± 1.8 [3.69] | |

\*\* No calibrated photometry is available for Keck M81 images, therefore, only relative values are given.

(1) The given aperture diameters yielded 68%, 90% and 95% of the integrated light of our average standard star profile.

(2) See Appendix for conversion of SiC magnitudes to 10.2 μm flux density.

(3) Extended Flux Ratio = $F_{ap}(r) / F_{r68,PSF}(r)$ is intended as an estimate of the extended flux compared to a point source flux normalized at the 68% radius measurement, $r_{68}$. $F_{ap}(r)$ is the aperture flux with radius r, $F_{r68,PSF}(r)$ = the flux of a point source with the given 68% radius aperture flux in an aperture of radius r, explicitly, $F_{r68,PSF}(r) = F_{ap}(r_{68}) PSF(r) / PSF(r_{68})$ where PSF(r) is the normalized integrated point spread function.

(4) The center of this aperture is 14 pixels from the edge of the image, therefore, an edge section of the circular aperture 3 pixels wide is missing from this measurement.

[R01] Roberts, Schurch, and Warwick (2001) give $L_X$ (0.5-10keV) = 39.53 and Γ=1.6; this 2-10keV value was derived assuming a perfect power law with the given index.

Table 2b Palomar/MIRLIN 2001 Imaging Results Summary

| | | Aperture Diameter[1] N (mag) | | | Aperture Diameter[1] Flux (mJy) [ ] | | | |
|---|---|---|---|---|---|---|---|---|
| Target (spectral class) | Log Lx (erg s$^{-1}$ cm$^{-2}$, 2-10 keV) | 0.8" 5.33 pix | 1.76" 11.73 pix | 2.72" 18.1 pix | 0.8" 10 pix | 1.76" 22 pix | 2.72" 34 pix | Time sou (tot |
| M81 N3031 ("LINER w/ marginal SY1.5") | 40.2 | | | | | | | |
| M87, NGC4486 | 40.70 | 7.25 ± 0.05 | 6.63 ± 0.06 | … | | | | |
| M58, NGC4579 | 41.18 | | | | | | | |

The MIRLIN instrument at Palomar has 0.15" square pixels, and a FWHM of 0.7" for these measurements.

Table 3 FWHM Values

| Target (spectral class) | FWHM (") maj. min. (pc at source) | FW90% (") maj. min. (pc at source) | d(Mpc) |
|---|---|---|---|
| M81 N3031 "LINER w/ marginal SY1.5" | 0.78 (13.7) 0.48 (8.5) | 1.7 (30) 1.02 (18) | 3.6 |
| NGC4203 L1.9 | 0.50 (37) 0.41 (30) | 1.4 (106) 0.95 (69) | 15.1 |
| M104[3], N4594 L2 | 0.63 (30) 0.49 (23) | 1.3 (62) 1.7 (81) | 9.8 |
| M94[3], N4736 L2 | 0.58 (12) 0.53 (11) | 1.6 (33) 1.9 (41) | 4.3 |

Note: These measures of the image profiles are meant as a rough characterization only. FWHM and major and minor axis measures are most appropriate when an object is known to have an elliptical-shaped profile, not obviously the case here. In order to provide such a measure, we examined the contour plots and chose an average "major axis" from the average maximum elongation axis for contours well above the noise but outside 1 FWHM. The minor axis was taken to be +90 degrees away. The result of this process is dependent on the contour algorithm, the noise in the image, and the smoothing applied. The cuts in the figures correspond to these major and minor axes, and are indicated on the contour plots.

Table 4 X vs. MIR Plot Data

| Object | D (Mpc) | 2-10 keV X-ray flux ($10^{-9}$ erg cm$^{-2}$ s$^{-1}$) | 10.2 μm MIR Flux (mJy) | Ref |
|---|---|---|---|---|
| M81 N3031 | | 143 | 161 | Grossan01 |
| NGC4203 | | 4.7 | 33 | This Work |
| M87, NGC4486 | | 4.9 | 9.2 | Telesco 2001 Palomar |
| M58, NGC4579 | | 45 | 66 | 2001 Palomar |
| M104[3], N4594 | | 12.1 | 30 | This Work |
| M94[3], N4736 L2 | | 10.4 | 59 | This Work |

**FIGURES**

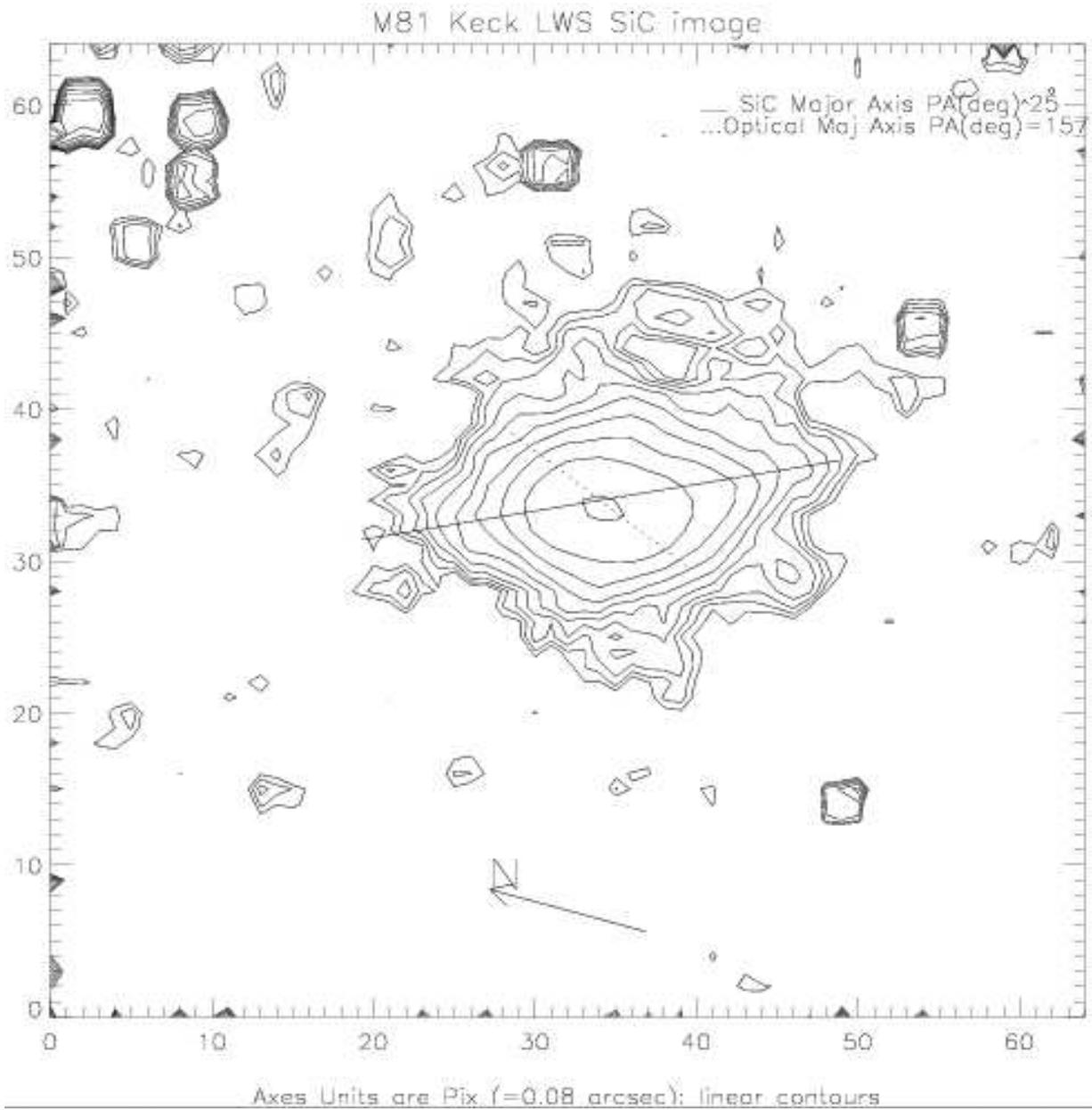

Figure 1 - SiC contour image of M81. The sold line shows an approximate average major axis (the minor axis, not shown, is chosen perpendicular). The dotted line shows the major axis from the RC3.

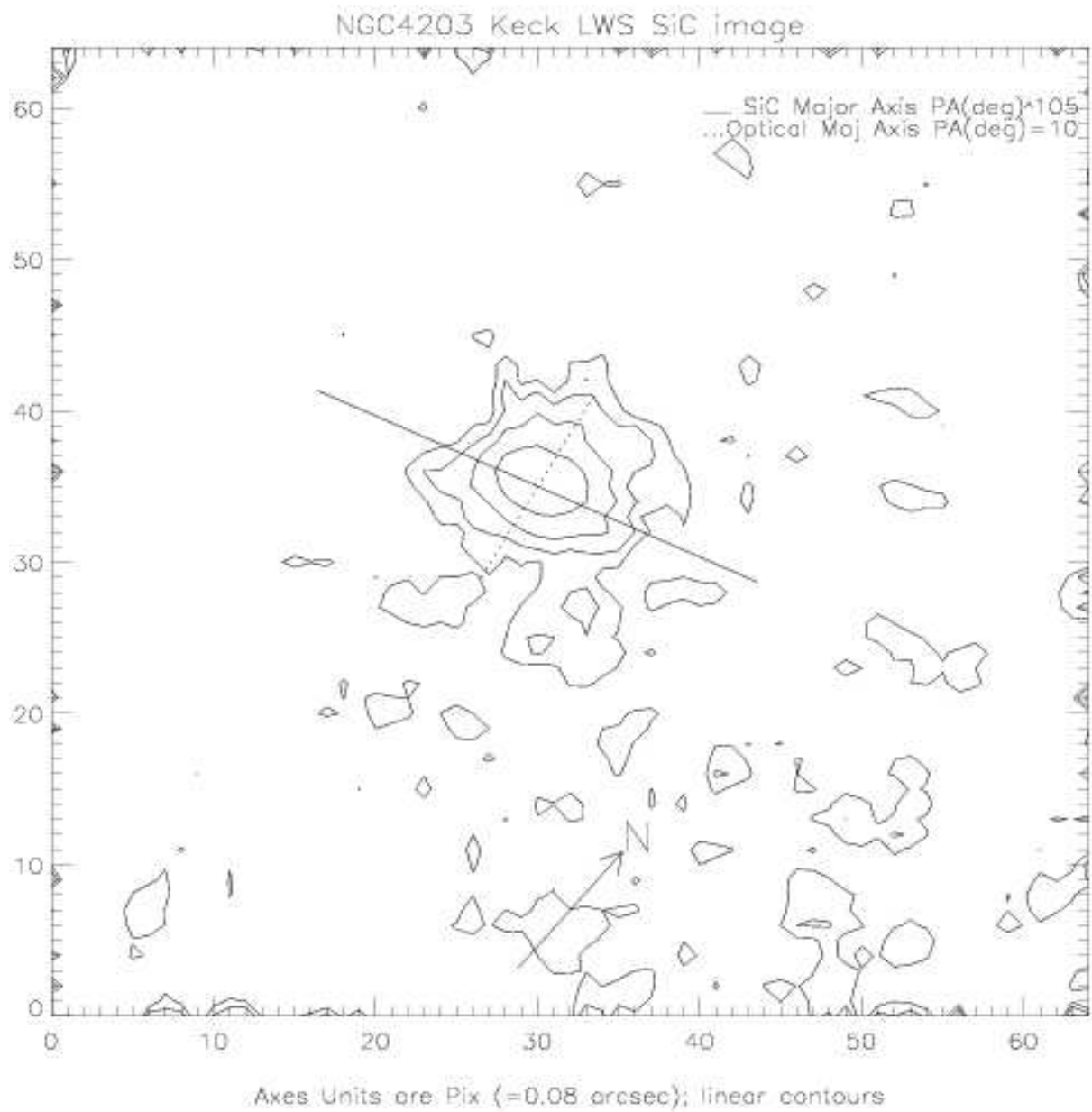

Figure 1 SiC image of NGC4203 (2003)

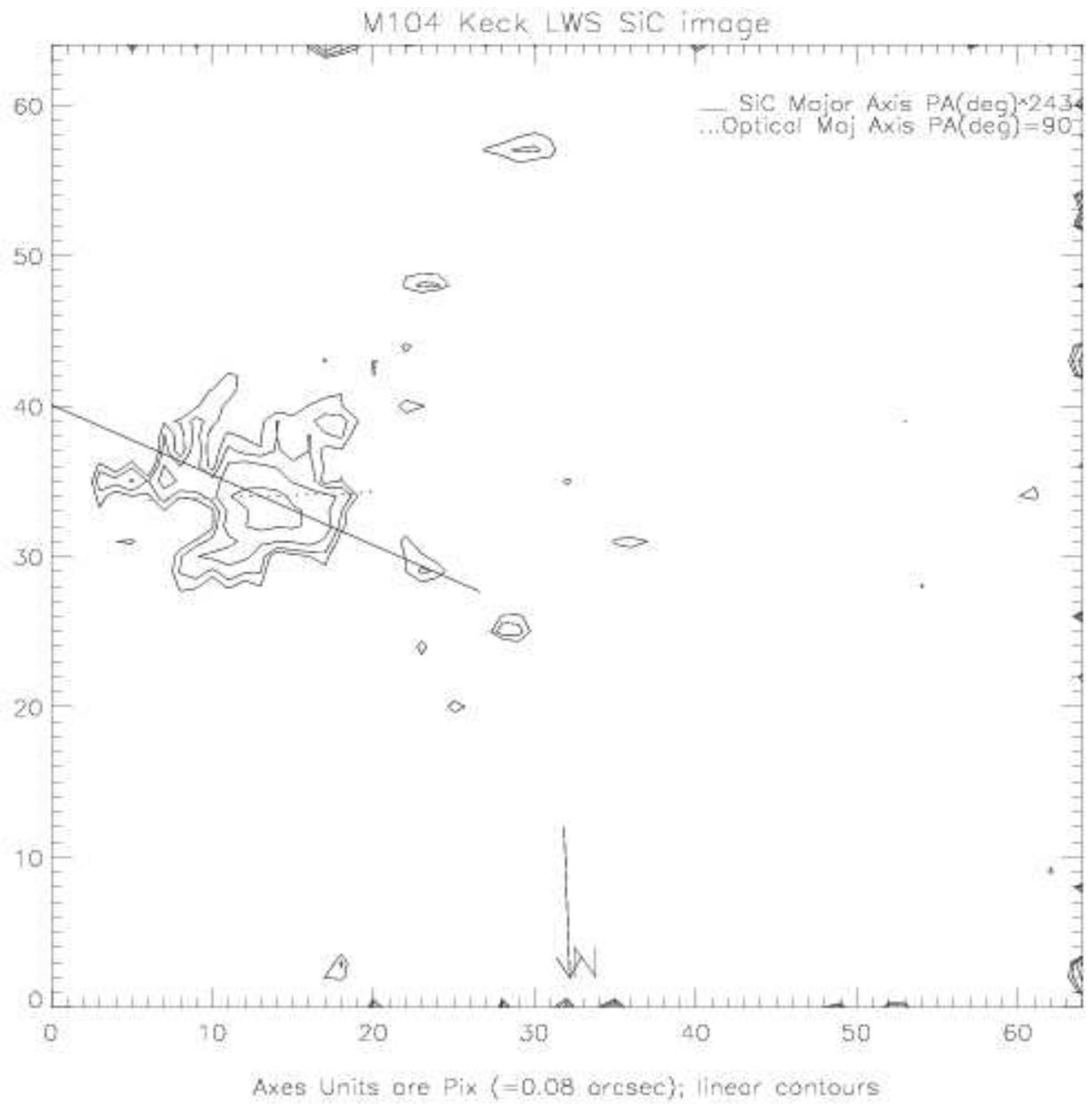

Fig. 1 SiC image of M104.

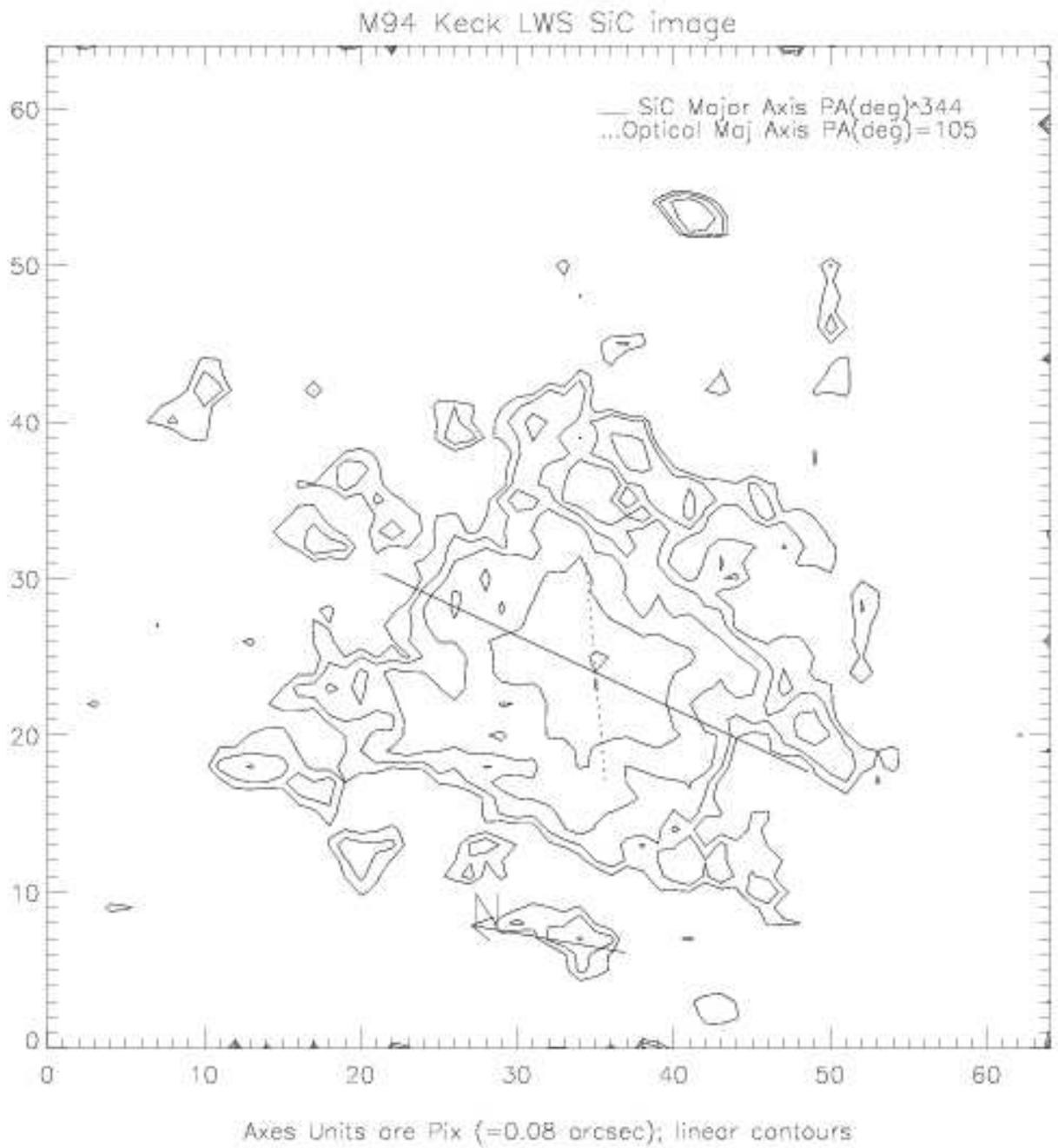

Fig 1. SiC image of NGC4736

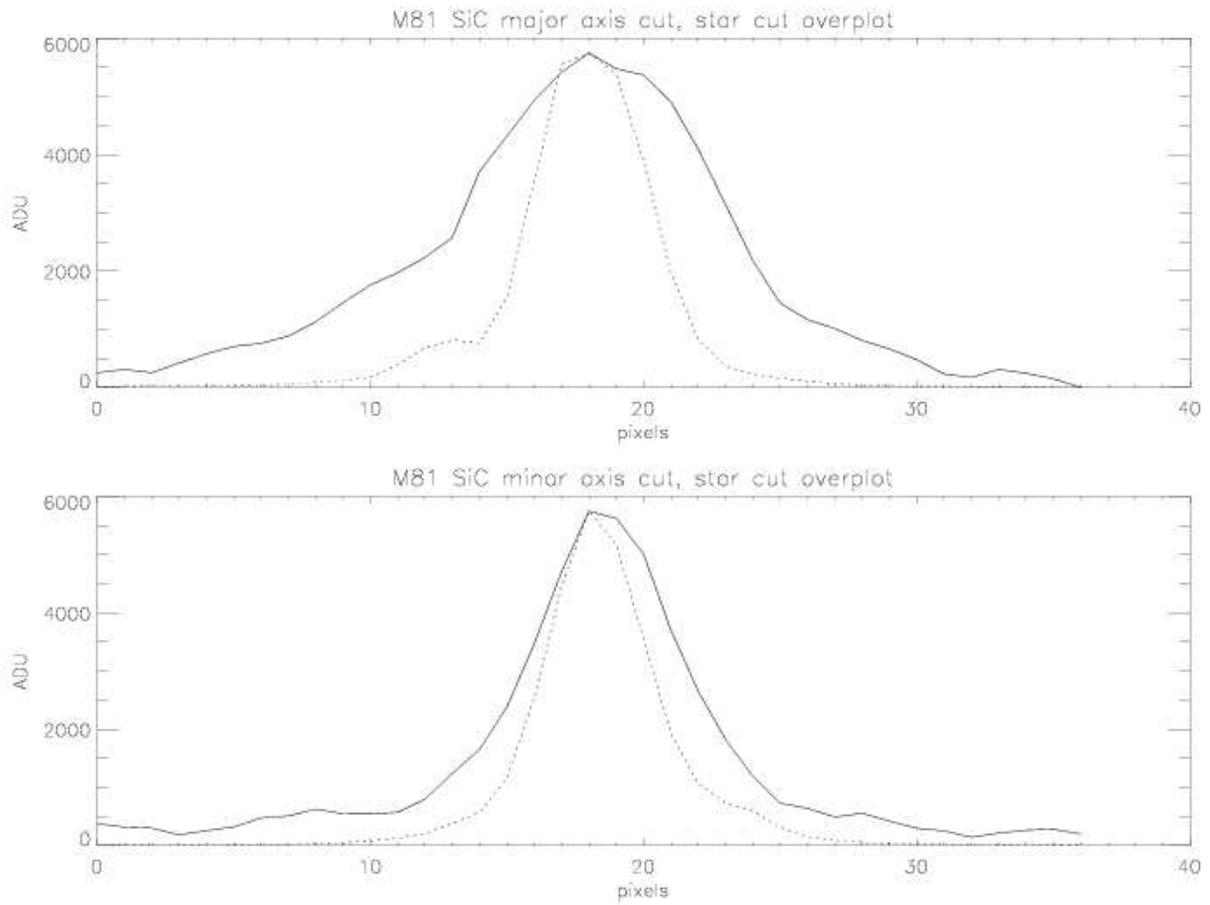

Figure 2a. Major and Minor cuts through the M81 profile are shown above. The axes are those shown on the contour images. The average PSF is represented by a dotted line. PSF cuts are taken along the same axes as the galaxy profile.

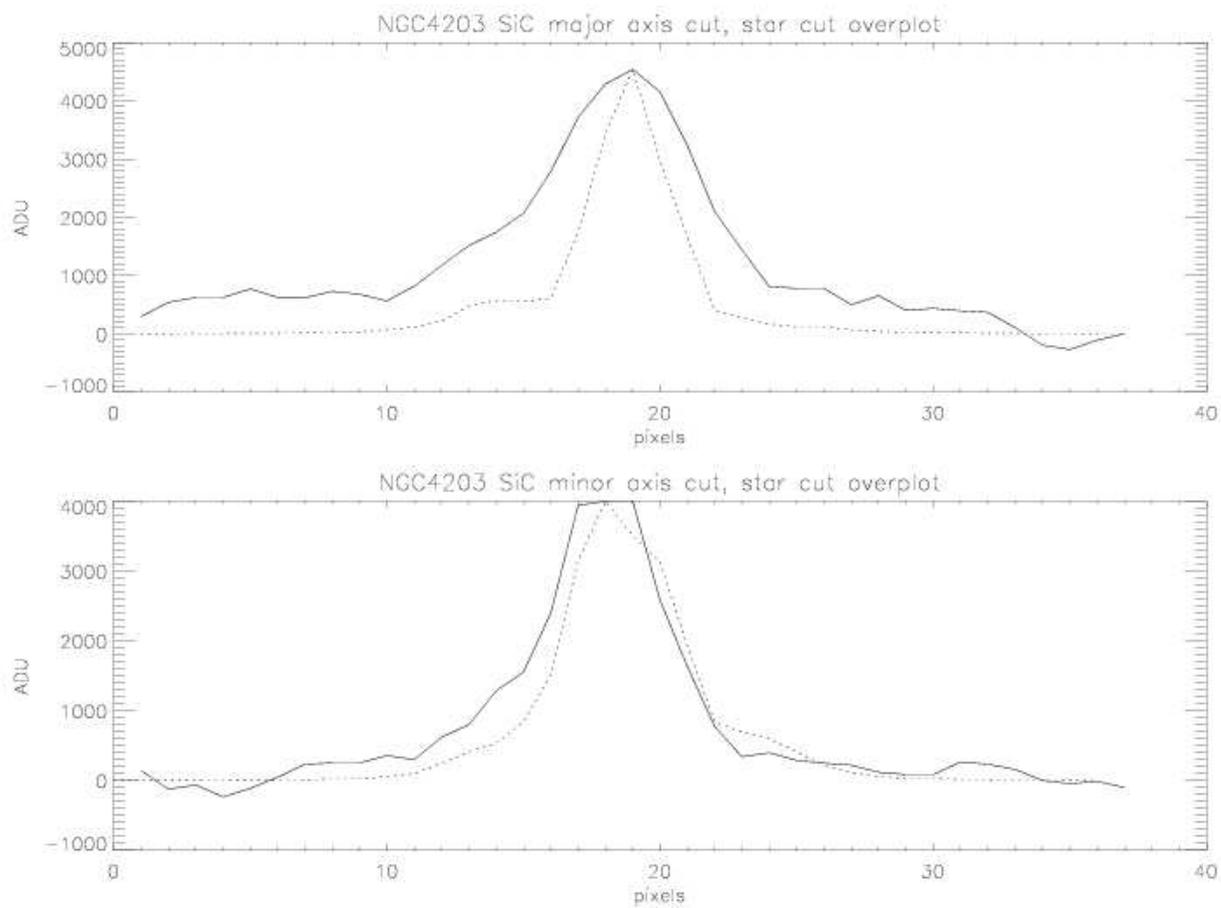

Figure 2b Major and Minor cuts through the NGC4203 profile.

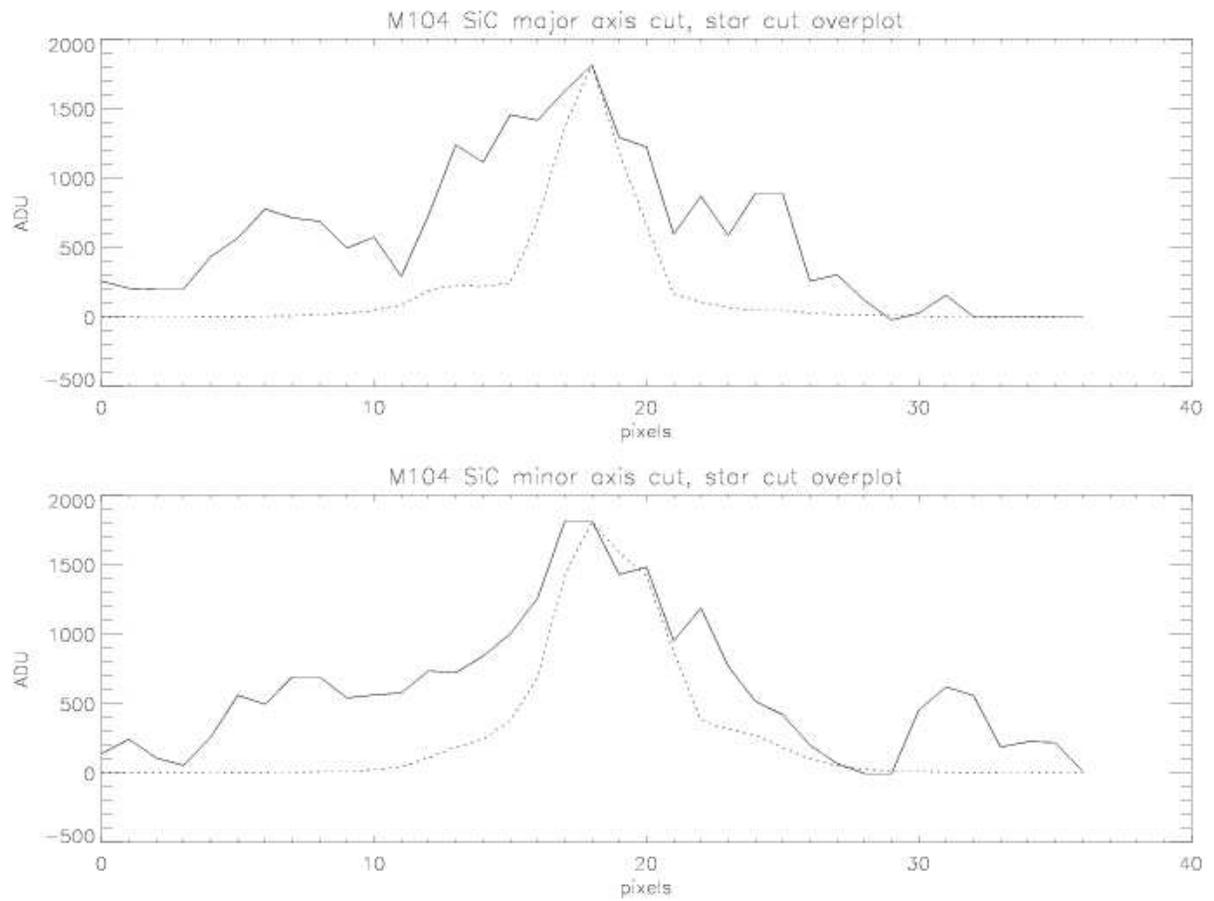

Figure 2c M104 major and minor axis cuts with PSF (dotted) for comparison.

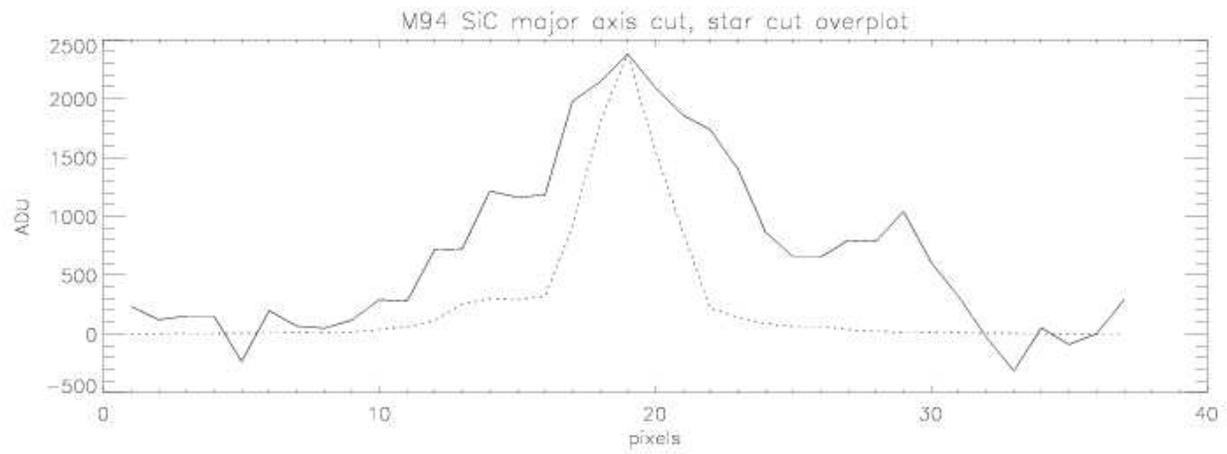
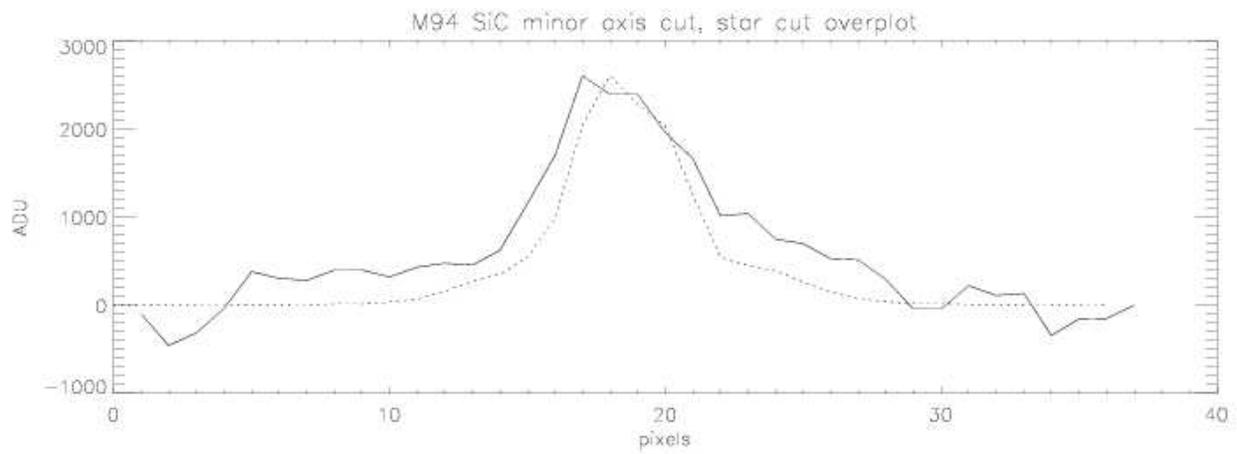
M94 (NGC 4736) major and minor axis cuts Th PSF (dotted) for comparison.

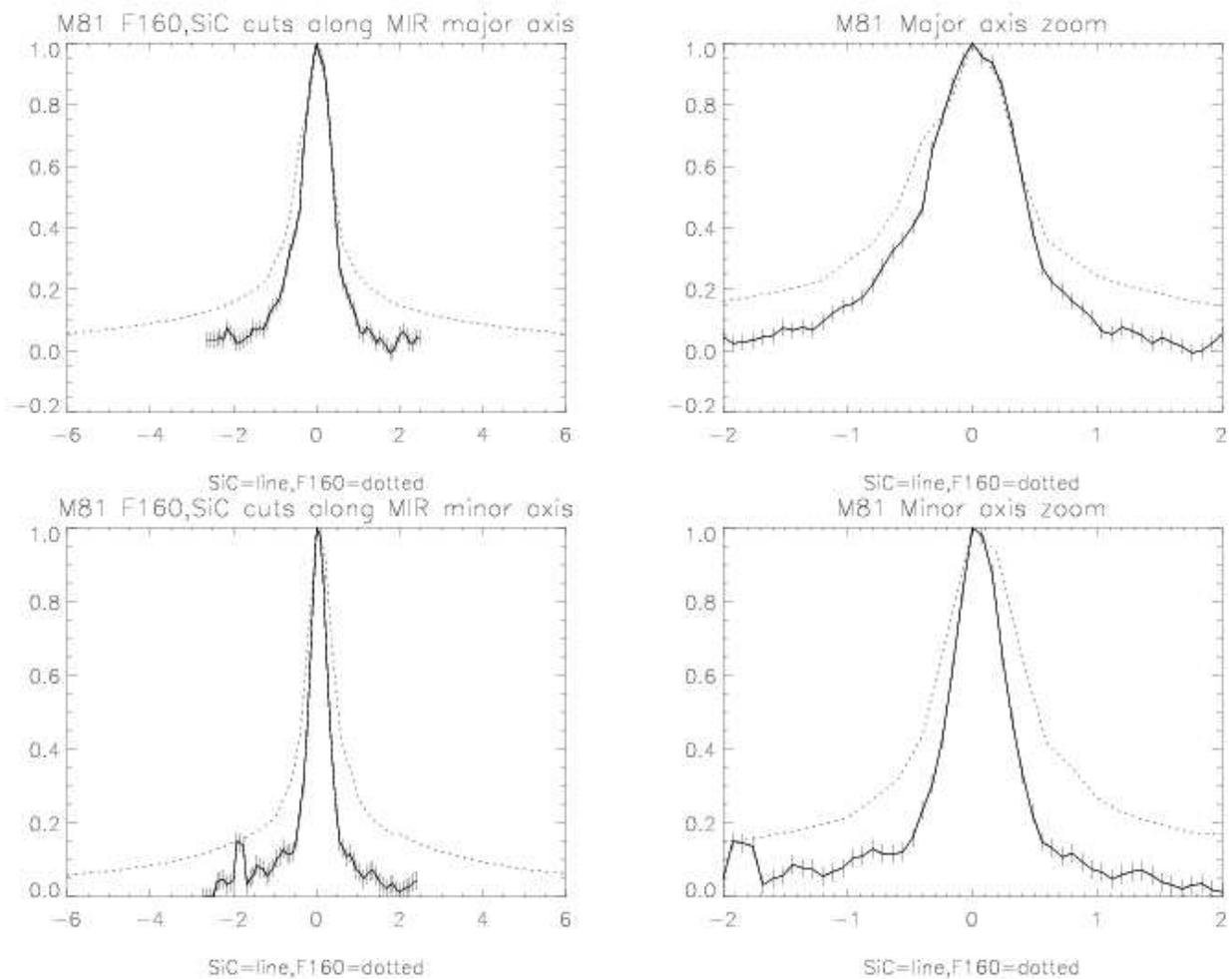

Figure 3. MIR major and minor axis cuts. The SiC and HST F160 profiles are shown together for comparison. Top figures show cuts along the MIR major axis, bottom figures show the mnor axis cuts. Small circles indicate the F160 profile, solid line with larger error bars indicates the SiC profile. The same figures are shown at expanded scale at right to emphasize the profile close to the nucleus. In all cases the MIR profiles are quite different from the NIR profiles.

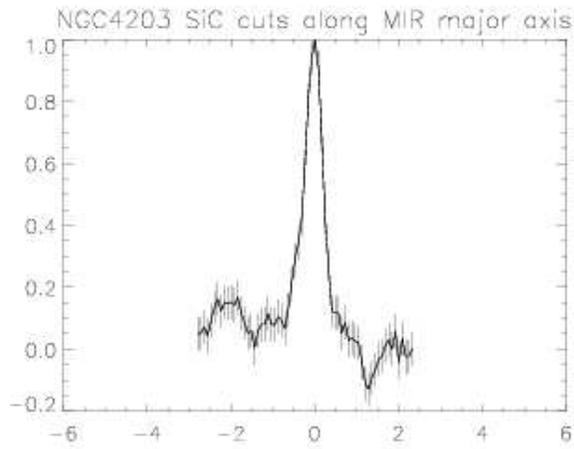 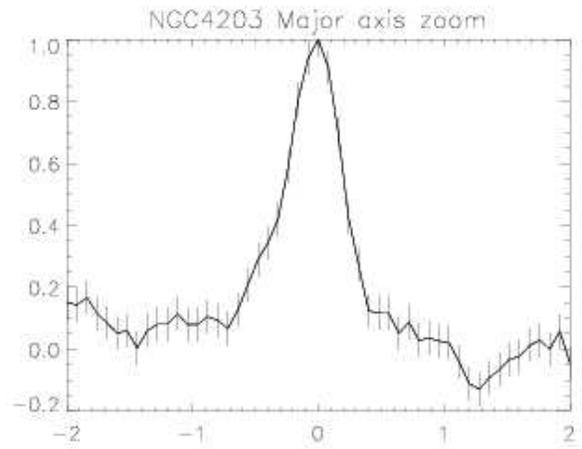
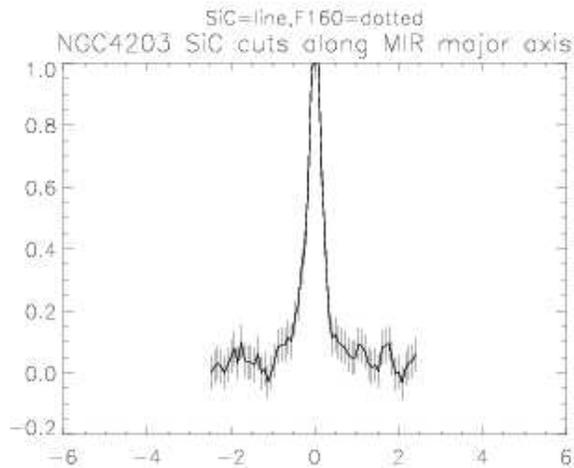 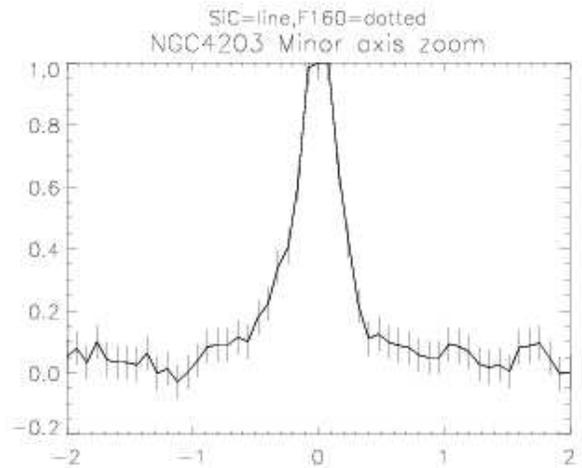

NGC4203 profile, shown same as other Keck LLAGN for comparison. At the time of publication, no NICMOS images of NGC4203 were avilable in the MAST archive, and we were unable to obtain NIR data of sufficiently high resolution from any other source.

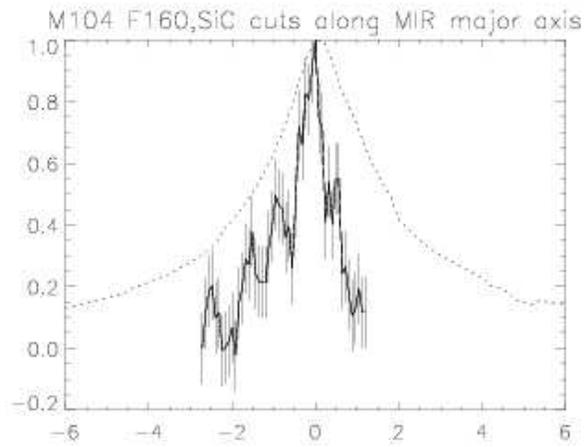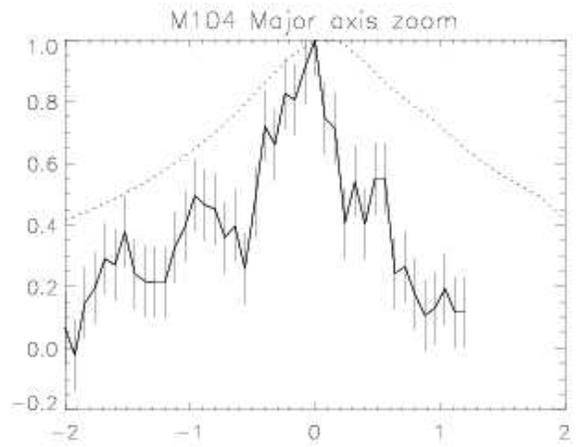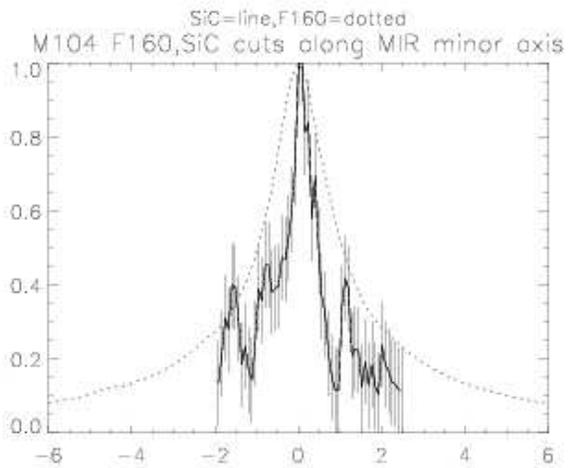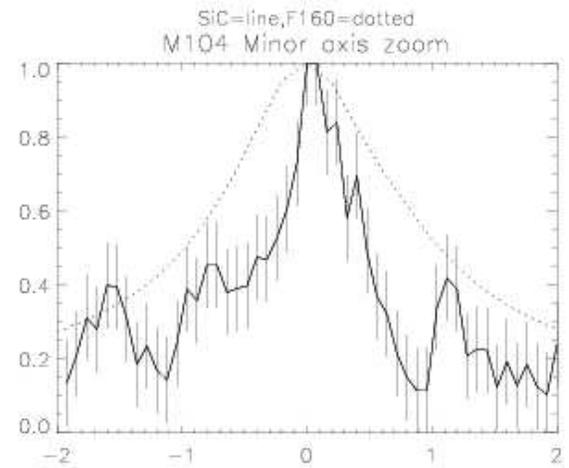

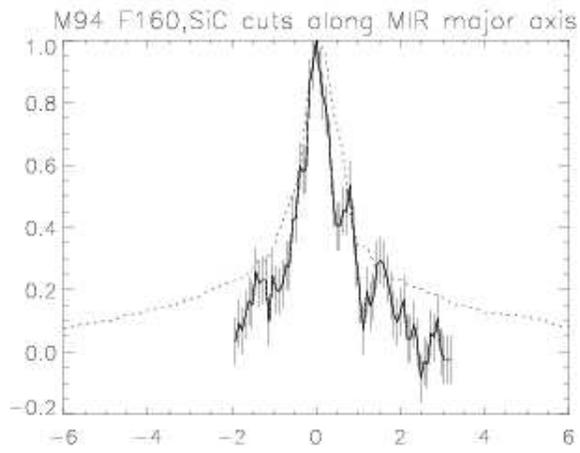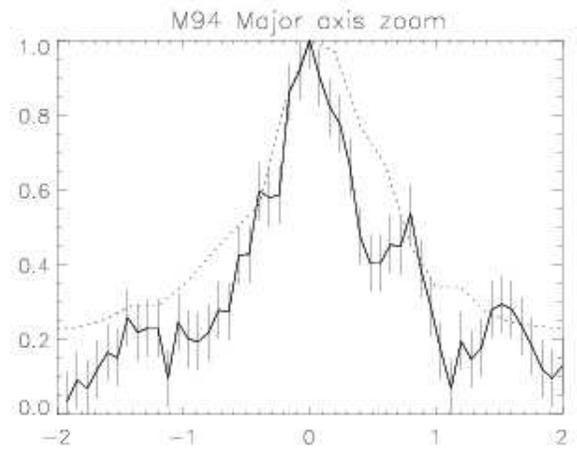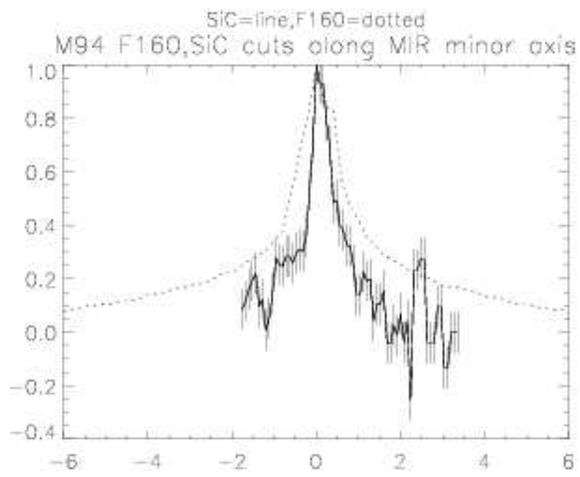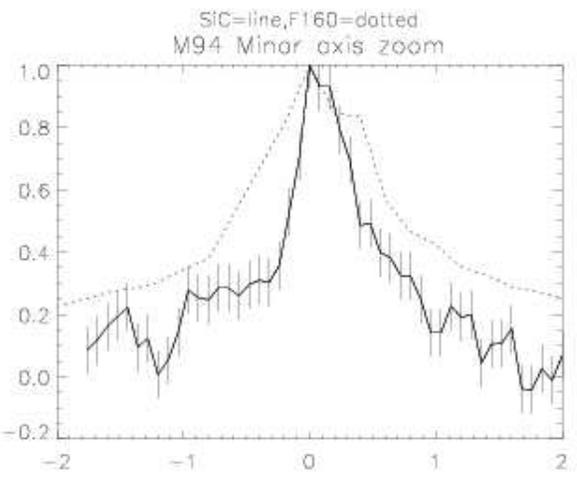

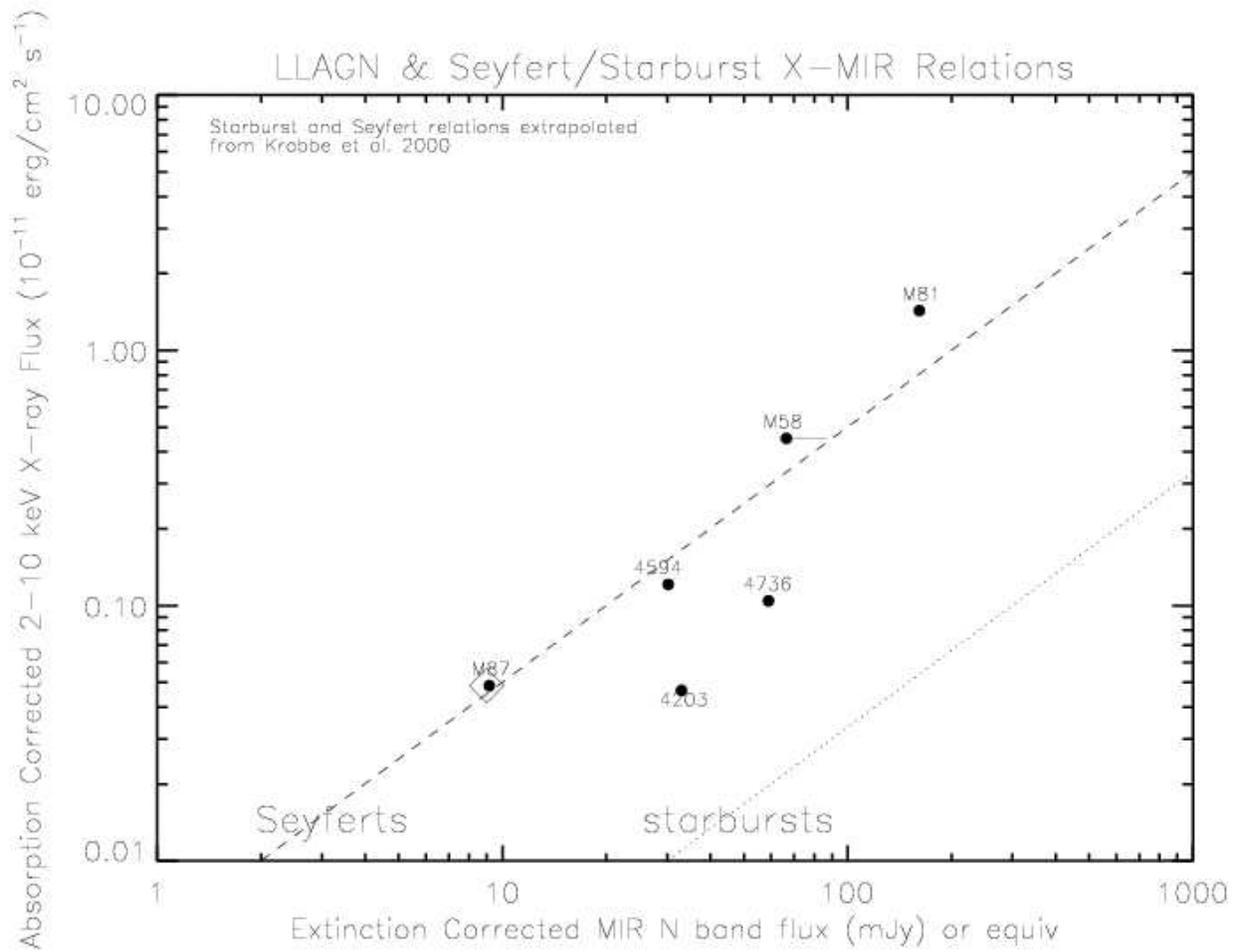

Figure 4a. X-ray flux vs. MIR flux plot for our sample. The Seyfert and Starburst lines are from KBF. The correlation holds well for all objects, with good separation from starbursts. See appendix for explanation of conversion of SiC fluxes to N band values. M87 is shown as with a special symbol to draw attention to the fact that it has a well-known radio jet which is detected in both MIR and X. Horizontal lines from the data points indicate uncertainties in absorption correction.

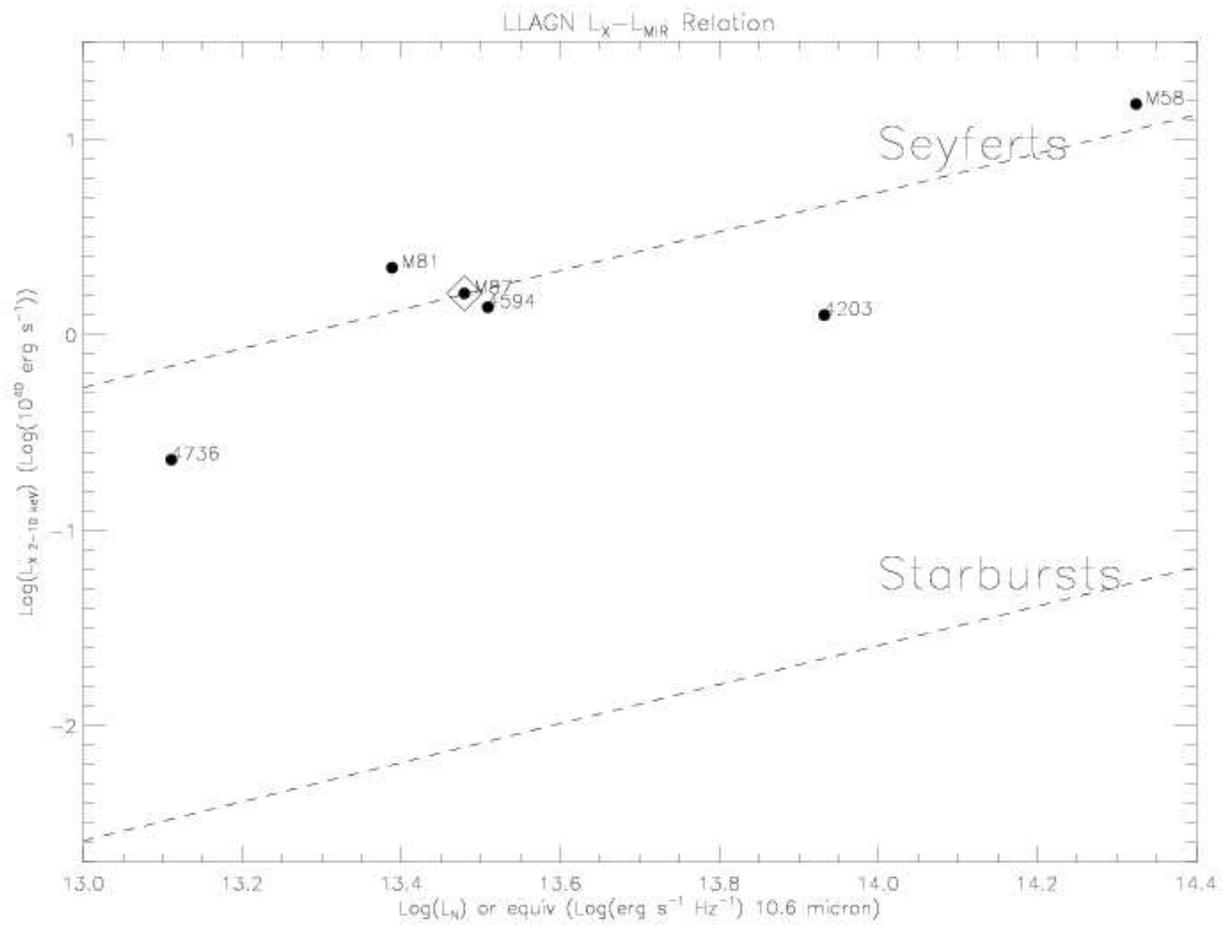

Figure 4b. $L_x$ vs. $L_{MIR}$ for our sample. The correlation clearly holds in luminosity.